# Investigation of the surface properties of different highly aligned N-MWCNT carpets

V. Eckert[a,*], E. Haubold[a], S. Oswald[a], S. Michel[b], C. Bellmann[b], P. Potapov[a,c], D. Wolf[a], S. Hampel[a], B. Büchner[a], M. Mertig[d,e], A. Leonhardt[a]

[a] IFW Dresden, Helmholtzstraße 20, 01069 Dresden, Germany

[b] IPF Dresden, Hohe Straße 6, 01069 Dresden, Germany

[c] Institut für Festkörper- und Materialphysik, Technische Universität Dresden, 01062 Dresden, Germany

[d] Institut für Physikalische Chemie, Technische Universität Dresden, 01062 Dresden, Germany

[e] Kurt-Schwabe-Institut für Mess- und Sensortechnik e.V. Meinsberg, 04736 Waldheim, Germany

**Abstract**

We investigated the physicochemical surface properties of different highly aligned nitrogen-doped multi-walled carbon nanotube (N-MWCNT) carpets, synthesized using toluene/pyrazine, toluene/benzylamine and acetonitrile via a sublimation-based chemical vapor deposition (SCVD) method at 760 °C. The surfaces of the N-MWCNT carpets synthesized using toluene/pyrazine and toluene/benzylamine were very hydrophobic. In contrast, we observed a complete wetting of the N-MWCNT carpets synthesized using acetonitrile. The difference in the wetting behavior of these N-MWCNT carpets is the main focus in this study and was not investigated before. Here, we show that not only the presence or concentration of nitrogen inside the carbon lattice, but especially it's kind of incorporation havean important influence on the surface polarity.

## 1. Introduction

Due to their extraordinary physical and chemical properties [1-3], pristine MWCNT promise a lot of applications such as (bio)sensors, catalyst support or as reinforcement for composite materials [4-8] and many others. The doping of such pristine MWCNT like e.g. with nitrogen leads to a change of the CNT morphology and properties. Basically, the incorporation of nitrogen into the carbon lattice causes structural defects, e.g. such as structural gaps or the presence of 5- or 7-member-rings due to pyridinic-bonded and pyrrolic-bonded nitrogen [9-11], which lead to more curvatures in the CNT structure [1]. But nitrogen doping can also lead to a straight CNT morphology under specific conditions [9,12-13]. Apart from the morphology changing, nitrogen doping leads to an increased electrical conductivity of the CNT, making them promising candidates for field emission tips, lithium storage and other nanoelectronic devices [14-15]. The doping process is commonly done *in situ*, which can be easily realized using the CVD method [16]. Moreover, nitrogen doping can also change the surface properties [17-20], and thus, enhances the dispersibility of CNT in water-based solutions. In general, pristine MWCNT exhibit a hydrophobic surface [17]. But when doped with nitrogen, the CNT surface has, compared to the undoped ones, more chemically activesites, which facilitate electron transfer and enhance a surface polarity of the CNT,and thus, a higher wettability [14,17]. This is especially known for MWCNT synthesized using acetonitrile and pyridine, which exhibit hydrophilic surfaces and can be easily dispersed in water-based solutions [17-18]. But it has been also reported that the surfaces of N-MWCNT synthesized using acetonitrile can be hydrophobic [21]. Further, a hydrophobic CNT surface can be also obtained when using the nitrogen-containing precursor benzylamine [22].

---

[*] corresponding author: v.eckert@ifw-dresden.de (Victoria Eckert)

A high wettability of MWCNT is generally very important for most applications in research and industry. Especially in applications using composite materials, e.g. CNT embedded in cement-based matrices to get a higher mechanical stability [8], a high dispersibility in the matrix material is needed. It is also favorable in biomedical investigations [23], such as *in vitro* tests, to reliably analyze the toxicological properties of individual MWCNT and not of MWCNT agglomerates.

Until to now, only undoped and a few nitrogen-doped MWCNT, as carpets or made of buckypapers [17, 24], were analyzed in regard of their surface properties. But detailed explanations regarding the differences in the surface properties between different N-MWCNT are still missing.

In our work we synthesized highly aligned N-MWCNT carpets using different nitrogen-containing precursors – pyrazine, benzylamine and acetonitrile – at 760 °C via the SCVD method and investigated their surface properties using dynamic water contact angle and X-ray photoelectron spectroscopy (XPS) measurements. Additionally, density-functional-theory (DFT) calculations of the electrical potential of the tube surfaces were carried out using the Vienna *ab initio* simulation package (VASP) to support the experimental results. We show that the surface properties strongly depend on the kind of nitrogen incorporation into the CNT structure rather than on the nitrogen concentration. Moreover, we emphasize the role of the constitution and decomposition of different nitrogen-containing precursors, which influence the nitrogen incorporation and concentration.

## 2. Experimental details

### 2.1 Synthesis of N-MWCNT

The N-MWCNT were synthesized using the SCVD method and a setup described previously [25]. We used ferrocene (purity ≥ 99.5 %, Alfa Aesar, CAS: 102-54-5) as the iron catalyst precursor, toluene (VWR Chemicals, CAS 108-88-3) as the pure C-precursor and pyrazine (purity ≥ 99.0 %, Merck KGaA, CAS: 290-37-9), benzylamine (purity ≥ 99.0 %, Merck KGaA, CAS: 100-46-9) as well as acetonitrile (purity ≥ 99.5 %, Merck KGaA, CAS: 75-05-8) as the nitrogen-containing precursors. Thereby pyrazine and benzylamine (each of them 30 wt.-%) were dissolved in toluene. To obtain the same carpet heights, the N-MWCNT were synthesized for different time periods. The reaction time was 20 min for toluene/pyrazine, 40 min for toluene/benzylamine and 15 min for acetonitrile. The reaction temperature was set to 760 °C. During the synthesis the N-MWCNT were deposited on $Si/SiO_2$ substrates (1 cm x 1 cm). The primary ferrocene amount was 1.9 g for each synthesis.

### 2.2 Characterization

The N-MWCNT carpets were analyzed using scanning electron microscopy (SEM; NanoSEM 200, 10 kV, FEI). Transmission electron microscopy (TEM; Tecnai T20, 200 kV, FEI) was used to characterize the MWCNT morphology. Atomic force microscopy (AFM, Dimension 3100, Fa. Veeco) in non-contact mode was used to analyze the surface roughness of the different N-MWCNT carpets. Dynamic water contact angle measurements were done using an OCA 35L (Dataphysics, Filderstadt). Ultrapure water ($\gamma_{LV}$ = 72.5 mJ/m$^2$) was used as the measuring liquid. Advancing and receding contact angles were directly determined on the as-synthesized N-MWCNT carpets. Therefore a liquid drop was increased up to 30 µl with a rate of 0.25 µl/s and withdrawn with the same speed. Three samples from each respective N-MWCNT type were measured. The contact angles were calculated using the tangent method of 5$^{th}$ order. It should be noted that the evaluation of the wettability of the different N-MWCNT was done more qualitatively because of their varying carpet heights and surface roughnesses which could affect the spreading behavior of the water drops. To analyze the concentration and the bond states of nitrogen and oxygen, XPS measurements were carried out (PHI 5600-Cl, Physical Electronics, 1486.6 eV, Al K$_\alpha$ radiation (350 W), 29 eV pass energy). DFT calculations were done using VASP 5.4.4. We simulated the charge densities of three different CNT models, all based on a triple-walled carbon nanotube (TWCNT) (5,0)@(14,0)@(23,0) with an infinite length, without affecting surrounding tubes. The structure was generated using VESTA (Visualization for Electronic and Structural Analysis, [26]) and contains 252 carbon atoms. To simulate doping with nitrogen, we added either 4 $N_2$ molecules or 6 $N_2$ molecules intercalated between the tube walls of the undoped system. This resulted in doping levels of 0.0 at.-% N, 3.0 at.-% N or 4.5 at.-% N for the calculated tubes. For the calculations of the charge densities of undoped and nitrogen-doped tubes, PBE (Perdew-Burke-Ernzerhof) pseudopotentials and a gamma-centered 5x5x5 monckhorstk-point set were used. The cutoff energy was set at 400 eV. To calculate the adsorption energy of water located close to surfaces of undoped and various N-doped TWCNT models, a water molecule was located with a distance of ~ 3.3 Å to the tube surface to avoid an overlap of the electron density between the TWCNT and the water molecule.

## 3. Results and discussion

### 3.1 *Surface roughness of N-MWCNT carpets*

Since we expect an influence of the CNT surface roughness on their wettability, we analyzed this surface roughness of the N-MWCNT carpets, which were on average about 25 µm high (Tab. 1 and Fig. 1 (a-c)) using AFM. But the AFM measurements were unsound and not reproducible, since the CNT carpets were not really suitable for this method. Because they strongly interacted with the cantilever, leading to stripes in pictures. The resulting surface roughnesses are shown in Tab. 1.

**Table 1**: Different N-MWCNT carpet heights and their surface roughnesses ($R_{rms}$ – root-mean-squared roughness; $R_a$ – arithmetical mean deviation of the assessed profile).

| N-MWCNT type synthesized using | Carpet height [µm] | Surface roughness [µm] | |
|---|---|---|---|
| | | $R_{rms}$ | $R_a$ |
| Toluene/benzylamine | 23.6 ± 3.6 | 0.412 | 0.568 |
| Toluene/pyrazine | 25.0 ± 1.9 | 0.054 | 0.260 |
| Acetonitrile | 25.0 ± 1.6 | 0.046 | 0.270 |

The N-MWCNT carpet synthesized using toluene/benzylamine show a high surface roughness, probably enhanced by a higher amount of amorphous carbon due to the much longer reaction time. In contrast, the N-MWCNT carpets synthesized using toluene/pyrazine and acetonitrile show less surface roughness, which are similar to each other. From our previous work, it is known that the toluene/pyrazine- and acetonitrile-MWCNT show mainly a tubular morphology, whereas the benzylamine-MWCNT exhibit a more bamboo-like morphology [25]. The reason for this morphology change was also discussed there. The morphology of this previous work is comparable with the morphology of the N-MWCNT in this work, synthesized at a slightly higher process temperature. Moreover, we suppose that the inner diameter of the tubes increases and the wall thickness decreases with increasing nitrogen concentration (Fig. 1, Tab. 2).

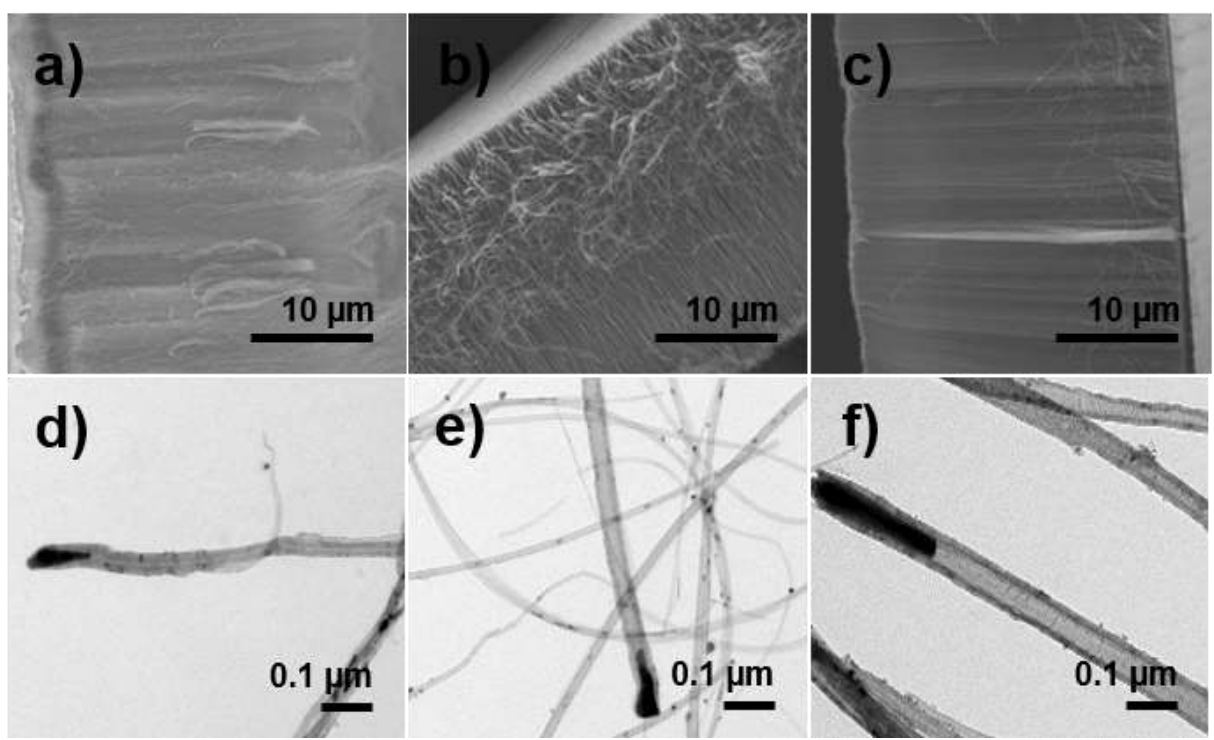

**Figure 1**: a)-c) SEM images of N-MWCNT carpets synthesized using toluene/benzylamine, toluene/pyrazine and acetonitrile (from left to right) and d)-f) corresponding TEM images of individual N-MWCNT after dispersion via sonication

The XPS measurement data given in Tab. 2 clearly show that acetonitrile provides the most nitrogen for the doping process compared to the other nitrogen-containing precursors, leading to the highest nitrogen content in the corresponding N-MWCNT.

**Table 2**: Atomic concentrations of nitrogen and oxygen indifferent precursors and the corresponding CNT product.

| N-containing precursor | N content [at.-%] Precursor* | N content [at.-%] N-MWCNT | O content [at.-%] N-MWCNT |
|---|---|---|---|
| Benzylamine | 1.88[a] | 0.35 | 0.61 |
| Pyrazine | 6.00[a] | 2.21 | 5.02 |
| Acetonitrile | 7.59 | 3.69 | 2.02 |

* calculated

[a] 30 wt.-% in toluene

The differences in the nitrogen content mainly depend on the constitution and the decomposition of the nitrogen-containing precursors. Normally, pyrazine itself provides a higher nitrogen concentration (20.00 at.-%) compared to acetonitrile, based on the thermal decomposition processes. Both, the direct and the decomposition in presence of $H_2$ lead to the formation of hydrogen cyanide and hydrocarbon, as seen in Equations 1, 1.1 for pyrazine and 2, 2.1 for acetonitrile [27-29]. But for our experiments, the pyrazine has to be mixed with toluene to generally prepare a reaction solution. Because of this, the initial nitrogen concentration was decreased to 6.00 at.-%.

$$C_4H_4N_2 \rightarrow C_2H_2 + 2HCN \qquad (1)$$

$$C_4H_4N_2 + H_2 \rightarrow C_2H_4 + 2HCN \qquad (1.1)$$

$$2CH_3CN \rightarrow C_2H_4 + 2HCN \qquad (2)$$

$$2CH_3CN + 2H_2 \rightarrow 2CH_4 + 2HCN \qquad (2.1)$$

In the case of benzylamine, the very low nitrogen concentration before and after its consumption, can be explained by the thermal decomposition process as seen in Equation 3 [30].

$$C_6H_5CH_2NH_2 + H_2 \rightarrow C_6H_5CH_3 + NH_3 \quad (3)$$

From Equation 3 it is clear that the nitrogen is present in form of ammonia after the thermal decomposition of benzylamine. We claim that the nitrogen incorporation during the CNT synthesis is easier if nitrogen is directly bonded to carbon like in case of hydrogen cyanide, as the decomposition product of pyrazine and acetonitrile. It can be suggested that ammonia is more stable than hydrogen cyanide at 760 °C. We assume that this is the main difference in the nitrogen content between toluene/benzylamine-MWCNT and toluene/pyrazine- as well as acetonitrile-MWCNT. Furthermore, it was discussed in [31] that acetonitrile also decomposes into active $CH_3\bullet$ and $CN\bullet$ radicals, whereas the $CN\bullet$ radicals are very suitable for nitrogen incorporation. The formation of active CN radicals during the thermal decomposition of pyrazine is also possible.

### 3.2 *Surface properties of N-MWCNT carpets*

The wettability of the respective N-MWCNT carpets was investigated using the dynamic water contact angle measurement method. The images of the sessile water drops on the surface of the N-MWCNT carpets and the corresponding curves of the advancing and receding contact angles are presented in Fig. 2, showing a clear difference in the surface properties between different doping levels. No spreading of the water drops on the surfaces of the N-MWCNT carpets synthesized using toluene/pyrazine and toluene/benzylamine was observed, because of their high advancing contact angles $\theta_A$, ~ 145°-150° and $\theta_A$~ 135°-145°, respectively. Their receding contact angles $\theta_R$ decreased from ~ 150° to ~ 125° for toluene/pyrazine and from ~ 145° to ~ 112° for toluene/benzylamine. During the receding, a pinning of the three-phase line and release was observed for both N-MWCNT types, as it can be seen in Fig. 2 (b, d). This effect can occur on chemically and/or morphologically heterogeneous surfaces, when i.e. additional interactions arise locally during the contact time between solid and liquid. Overall, the surfaces of these N-MWCNT carpets are very hydrophobic. A hydrophobic surface character was expected for the N-MWCNT carpets synthesized using toluene/benzylamine due to the very low nitrogen content, leading to the same behavior like an undoped sample. In contrast, a complete wetting was observed for the N-MWCNT carpets synthesized using acetonitrile. Their advancing contact angles $\theta_A$ were ~ 45° and the receding contact angle were less than $\theta_R$< 10°, showing hydrophilic surfaces (Fig. 2 (e-f)).

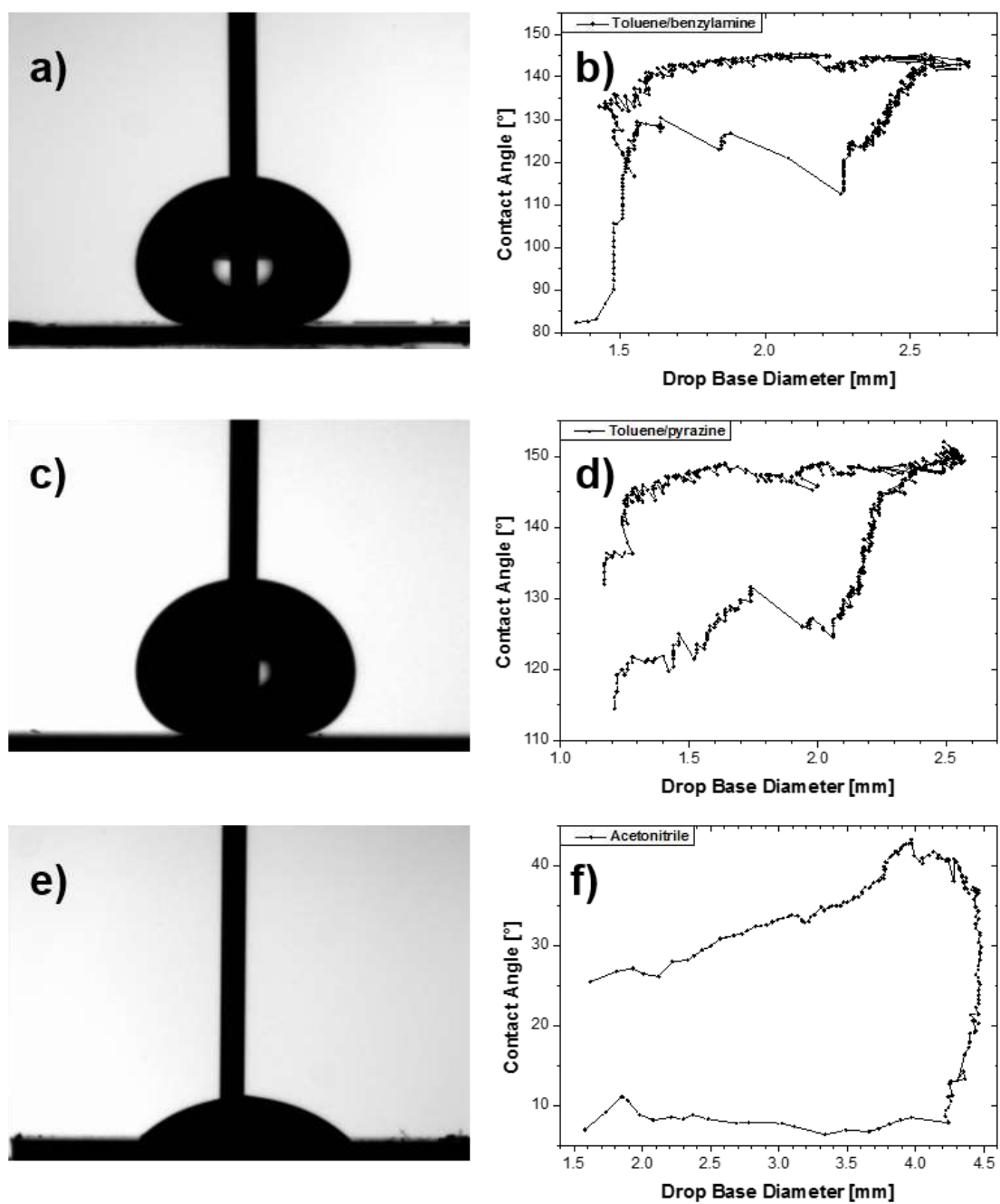

**Figure 2**: Images of the water droplets of the dynamic contact angle measurements (a, c, e) and their corresponding contact angle data (b, d, f). From top to bottom: toluene/benzylamine, toluene/pyrazine, acetonitrile.

As mentioned before, the surface roughness and also the nitrogen concentration can influence the water spreading behavior. However, we could not observe an influence of the surface roughness on the wettability of the N-MWCNT carpets. In particular, the surface roughnesses of the toluene/pyrazine- and acetonitrile-MWCNT carpets are very similar, as already mentioned, leading to different surface properties. There is no influence of the CNT morphology on the wettability. Moreover, we suggest that the oxygen concentration in Tab. 2 has no significant influence on the surface properties. In particular, the N-MWCNT carpets synthesized using toluene/pyrazine exhibit a high oxygen content compared to the other N-MWCNT carpets and show hydrophobic surfaces.

The different surface properties of the N-MWCNT carpets are likely influenced by the nitrogen concentration (Tab. 2) and the kind of its incorporation into the carbon lattice. The bonding configuration of carbon and nitrogen was obtained from high-resolution XPS spectra in the C1s and N1s regions, both fitted with Gaussian/Lorentzian function, and presented in Fig. 3.

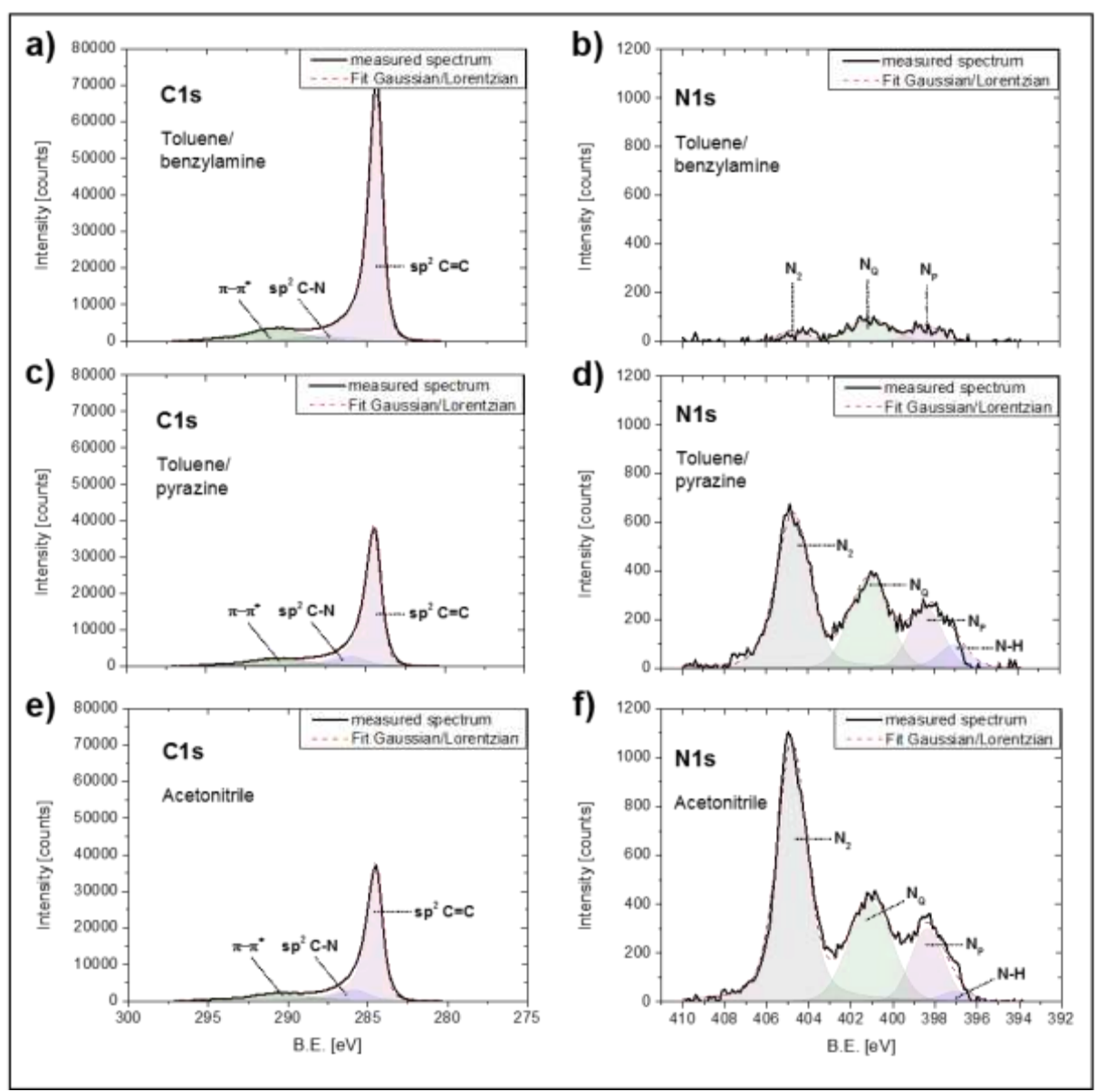

**Figure 3**: High-resolution XPS spectra in C1s (a, c, e) and N1s regions (b, d, f) of different N-MWCNT, fitted with Gaussian/Lorentzian function.

Applying the Gaussian function, we analyzed numerically the overlapping transitions through modelling of the line-shapes fitted to the data. The C1s spectra in Fig. 3 (a, c, e) can be deconvolved into three individual peaks, representing the $sp^2$-hybridized carbon peak at 284.4 eV, the $sp^2$-hybridized C-N peak at 285.8 eV and a peak at 290.4 eV resulting from $\pi$-$\pi^*$ electronic transitions in aromatic rings [17]. From Tab. 3 it can be seen that the acetonitrile-MWCNT exhibit a slightly higher amount of C-N configurations due to higher nitrogen doping compared to toluene/pyrazine-MWCNT. In case of the benzylamine-MWCNT, the concentration of the C-N configuration is clearly negligible because of the very low nitrogen doping.

**Table 3**: Concentrations of bonding configuration in C1s spectra of different N-MWCNT.

| N-MWCNT type synthesized using | Peak area (%) | | |
|---|---|---|---|
| | $sp^2$ C=C | $sp^2$ C-N | $\pi$-$\pi^*$ |
| Toluene/benzylamine | 84.4 | 3.3 | 12.3 |
| Toluene/pyrazine | 76.0 | 10.4 | 13.6 |
| Acetonitrile | 74.9 | 11.7 | 13.4 |

The N1s spectra of the respective N-MWCNT (Fig. 3 (b, d, f)) clearly show different nitrogen functionalities and can be deconvolved into three major peaks: the pyridinic nitrogen ($N_P$) at a binding energy of 398.3 eV – 398.8 eV, quaternary nitrogen ($N_Q$) at 400.9 eV – 401.5 eV and a peak at 404.6 eV – 405.0 eV [17]. In case of the toluene/pyrazine- and acetonitrile-MWCNT, the very small peak at 397.0 eV can be assigned to tetrahedral nitrogen bonded to $sp^3$-C. The reason for the presence of this peak could be an undecomposed N-H bond [1]. The nature of the peak at ~405.0 eV was often discussed in the literature. Choi and Park [32] suggested that the existence of this peak depends on a high photon energy of 1265 eV, leading to a higher photoelectron escape depth. They defined this peak as gaseous nitrogen, being intercalated between the CNT shells in the vicinity of an internal cavity of the CNT. In contrast, this peak can also be identified as chemisorbed $-NO_x$ species, like a nitro group $-NO_2$ or in form of nitrogen oxides of pyridinic nitrogen [33-36]. For the proof of nitrogen oxides, we measured the N1s and O1s spectra of freshly synthesized N-MWCNT carpets, which were exposed to air for 1 h. We again obtained the peak at around 405.0 eV, but no clear traces of oxygen (~0.1 at.-

%, as shown in supplementary data). As the oxygen concentration is too low, it is unlikely that the oxygen is bonded to nitrogen, but rather to carbon. So we can clearly conclude, this peak is related to molecular nitrogen ($N_2$) in the CNT core or intercalated between the tube walls. Some of these nitrogen functionalities are known to enhance the polarity of the tube surface. The pyridinic nitrogen acts as an electron donor because of their localized electron lone pair and exhibits a strong basicity. In contrast, the quaternary nitrogen acts as an electron acceptor due to its cationic structure and facilitates electron transfer [15, 17]. Both of them can change the local electron density of the tube walls, enhancing the surface polarity. In case of the benzylamine-MWCNT, the concentration of the pyridinic and quaternary nitrogen is very low (Tab. 4), because of the very low total nitrogen concentration (Tab. 2), which is insufficient to increase the polarity of the tube surface. The comparison between the toluene/pyrazine- and acetonitrile-MWCNT regarding to the amount of the pyridinic and quaternary nitrogen, does not show any significant difference (Tab. 4).

**Table 4**: Concentrations of N functionalities in N1s spectra of different N-MWCNT.

| N-MWCNT type synthesized using | Peak area (%) | | | |
| --- | --- | --- | --- | --- |
| | $sp^3$ C-N | $N_P$ | $N_Q$ | $N_2$ |
| Toluene/benzylamine | 0.0 | 26.6 | 54.7 | 18.6 |
| Toluene/pyrazine | 6.4 | 16.1 | 28.3 | 49.2 |
| Acetonitrile | 2.0 | 14.4 | 26.2 | 57.4 |

But from Fig. 2 it is known that the toluene/pyrazine-MWCNT are clearly hydrophobic compared to the very hydrophilic acetonitrile-MWCNT. In that case, the polar nitrogen functionalities at the surface of the toluene/pyrazine-MWCNT do not lead to an increased hydrophilicity. The only significant difference is seen in the intensity of the molecular nitrogen at around 405.0 eV, which is clearly higher for acetonitrile-MWCNT compared to toluene/pyrazine-MWCNT. As mentioned before, the molecular nitrogen can be intercalated between the tube walls, preferentially between the inner tube walls [37]. The probe depth is about 5 nm using a photon energy of 1486.6 eV and the total wall thickness of the acetonitrile- and toluene/pyrazine-MWCNT is about 15-20 nm and 18-25 nm. In that case, the molecular nitrogen intercalated between the inner tube walls cannot be detected. According to the high intensity of the molecular nitrogen, we assume that it is more distributed towards outer tube walls, without influencing the CNT interlayer distance. The averaged interlayer distances of the N-MWCNT and a further proof of the existence of molecular nitrogen using Electron Energy Loss Spectroscopy are shown in the supplementary data.

Even though molecular nitrogen itself is uncharged, its electron density can influence the electron density of the delocalized π electron system and the electrical potential. Due to the curvature of the CNT shells, each shell exhibits an electron-rich outer surface due to an electron density shift [38]. When a nitrogen molecule is intercalated between the tube walls, its electron density can cause a larger shift of the π electron density.

Since we know from XPS measurements that there is no significant difference in the amount of polar nitrogen functionalities between toluene/pyrazine- and acetonitrile-MWCNT, we only considered the amount of molecular nitrogen for further discussion regarding the different surface properties. To investigate a possible influence of intercalated nitrogen molecules on the tube polarity, we used TWCNT models with different amounts of molecular nitrogen, 0.0 at.-% N, 3.0 at.-% N and 4.5 at.-% N, to calculate charge densities using VASP, as shown in Fig. 4. Due to delocalized electrons, we only considered lower charge densities to analyze a possible shift of the electron density.

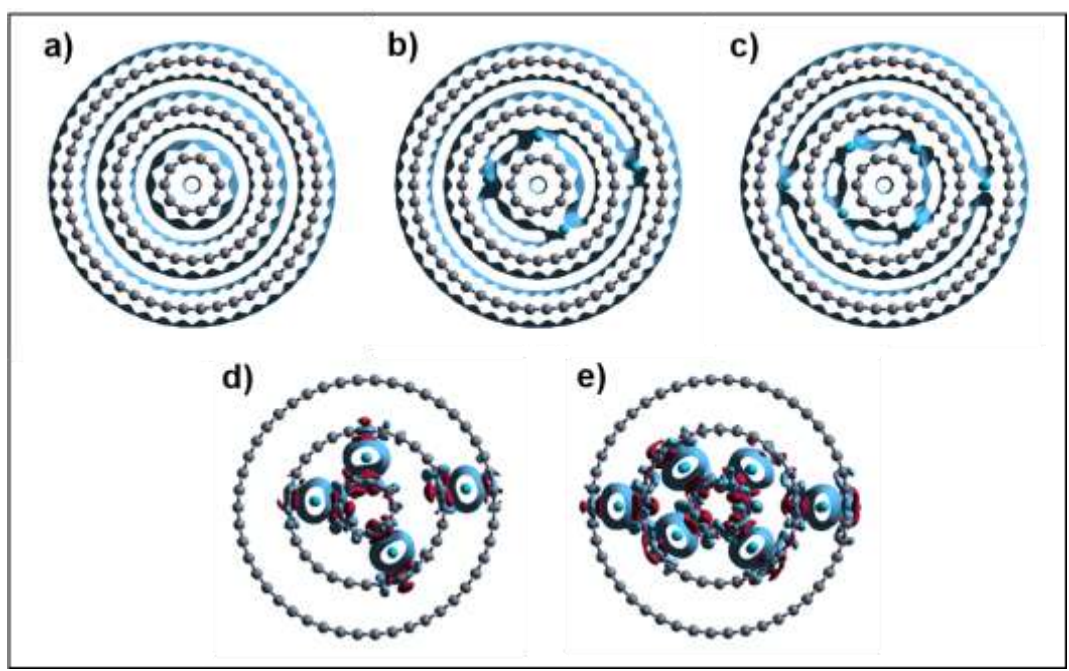

**Figure 4**: Isosurfaces of charge densities of a TWCNT model (5,0)@(14,0)@(23,0). a) undoped, b) low nitrogen-doped (3.0 at.-%), c) high nitrogen-doped (4.5 at.-%), d)-e) charge

densities of b) and c) after subtracting the undoped model a) (red: positive charge due to electron deficiency, blue: negative charge).

In Fig. 4 (a) we calculated the entire charge density of an undoped TWCNT model. The outer tube surface exhibits a higher electron density compared to the inner tube surface due to the curvature, which decreases with increasing tube diameter (decreasing curvature), being in agreement with [38]. The intercalation of molecular nitrogen between the tube walls (parallel orientation to CNT axis, as in [37]) in Fig. 4 (b-c) clearly induces a shift of the π electron density of the tube walls. The electron densities of the outer surface of the (5,0) tube and of the inner surface of the (14,0) tube were shifted towards the intercalated nitrogen molecule. This leads to an electron deficiency at the outer and inner tube surface. To discuss this electron density shift in more detail, the entire charge density of the undoped TWCNT model (Fig. 4, a) was subtracted from the charge densities of the nitrogen-doped TWCNT models (Fig. 4(b- c)), shown in Fig. 4 (d-e). As expected, the electron density shift towards the nitrogen molecule results in a higher electron density surrounding the nitrogen molecule (blue), leaving an electron deficiency (red) at the outer surface in case of the (5,0) tube and at the inner surface in case of the (14,0) tube. It clearly indicates that the intercalation of molecular nitrogen between the tube walls strongly influences the delocalized π electron system. For comparison, the influence of pyridinic and quaternary nitrogen on the π electron density of the TWCNT models is shown in the supplementary data.

To support the results of the dynamic contact angle measurements in Fig. 2, the adsorption energy of a water molecule located close to the surfaces of different TWCNT models was calculated, as shown in Tab. 5. To be adsorb, a negative value of the calculated energies is required.

**Table 5**: Calculated adsorption energies ($E_{ad}$**) of a water molecule located close to different surfaces of TWCNT models (distance: ~ 3.3 Å)

| TWCNT surface | Adsorption energy [eV] of $H_2O$ |
| --- | --- |
| Undoped | 0.013 |
| N-doped (4 $N_2$) | 0.004 |
| N-doped (6 $N_2$) | -0.444 |

** $E_{ad} = E_{tot}(CNT + water) - (E_{tot}(CNT) + E_{tot}(water))$ [40]

The undoped and various N-doped TWCNT models (additional pyridinic and quaternary nitrogen) with a water molecule located closed to their surfaces are shown in the supplementary data. When looking at the different results of the calculated TWCNT models, only the surface of a TWCNT model containing a higher concentration of intercalated molecular nitrogen would be able to adsorb the water molecule, fitting very well with the experimental observation. The lower amount of molecular nitrogen is not sufficient to adsorb the water molecule on the surfaceof the TWCNT model.

It should be taken into account that we calculated only TWCNT models, applying the results of the charge densities to our real N-MWCNT types with some more walls. The use of the TWCNT model is sufficient to support the interpretation, that a certain amount of intercalated molecular nitrogen is needed to change the surface properties of our synthesized N-MWCNT. If we reconsider the different surface properties between the toluene/pyrazine- and acetonitrile-MWCNT we can assume that the hydrophilicity increases especially with increasing intercalated molecular nitrogen. The high concentration of molecular nitrogen inside the acetonitrile-MWCNT suggest that these molecules are more distributed towards outer tube walls, enhancing a homogeneous hydrophilicity at the outer tube surface. Taking into account that the nitrogen molecules are intercalated first between the inner tube walls [37], increasing the amount of these molecules causes them to be intercalated between outer tube walls, as the space between the walls is limited. In the case of toluene/pyrazine-MWCNT, the much lower concentration of molecular nitrogen may indicate an insufficient distribution of these molecules towards outer tube walls, leading to a heterogeneous polarity of the outer tube surface. This results in a tube surface with a lot of hydrophobic parts and some hydrophilic parts, which is insufficient for a complete dispersion in water-based solutions. The influence of one nitrogen molecule intercalated between different tube walls is shown in the supplementary data.

## 4. Conclusions

We showed that different N-MWCNT exhibit different surface properties – from hydrophilicity to hydrophobicity. Doping with nitrogen does not always increase the surface polarity of the tubes. In this work, even a doping level at ~2.2 at.-% N, as in the case for toluene/pyrazine-MWCNT, did not enhance hydrophilicity of the CNT surface. In contrast, higher doped N-MWCNT (~3.7 at.-% N) synthesized using acetonitrile exhibit hydrophilic surfaces. We found, that especially the type of nitrogen functionality – incorporated in the carbon lattice as quaternary and pyridinic nitrogen or intercalated between the tube walls as molecular nitrogen – has a stronger influence on the surface properties than the nitrogen concentration.

Acetonitrile-MWCNT exhibit much more molecular nitrogen intercalated between their tube walls than quaternary and pyridinic nitrogen incorporated inside their carbon lattice, compared to toluene/pyrazine-MWCNT. The amount of polar nitrogen functionalities is nearly the same for both N-MWCNT types. The only significant difference between the hydrophobic toluene/pyrazine-MWCNT and the hydrophilic acetonitrile-MWCNT is the concentration of molecular nitrogen and not of the polar nitrogen functionalities. With the calculation of the charge densities, using appropriate CNT models, we highlighted the strong influence of the intercalated molecular nitrogen on the delocalized $\pi$ electron system of the tubes. Intercalation of molecular nitrogen induces a shift of the delocalized $\pi$ electron density towards the nitrogen molecule, leading to an electron deficiency at the outer tube (5,0) and inner tube (14,0) surfaces. The additional calculation of the adsorption energy of a water molecule located close to the surfaces of undoped and various N-doped TWCNT models reveals that the water molecule is only adsorbed on the surface of the TWCNT model containing a higher concentration of intercalated molecular nitrogen. Making a CNT surface hydrophilic, the molecular nitrogen has to be distributed between the outer tube walls in an appropriate concentration leading to a homogeneously polarity of the tube surface.

**Acknowledgments**

The authors thank the BMBF project CaNTser - Investigation of the toxic potential of carbon nanotubes after longtime-inhalation (FKZ: 03XP0016) for funding, Steffi Kaschube for the XPS measurements, Barbara Eichler for the AFM measurements and Ulrike Nitzsche for technical assistance in using VASP. Furthermore Daniel Wolf acknowledges funding from the European Research Council via the ERC-2016-STG starting grant atom. Pavel Potapov acknowledges funding from DFG "Zukunftskonzept" (F-003661-553-Ü6a-1020605).

**References**

[1] M. I. Ionescu, Y. Zhang, R. Li , H. Abou-Rachid, X. Sun. Nitrogen-doping effects on the growth structure and electrical performance of carbon nanotubes obtained by spray pyrolysis method. Appl. Surf. Sci. 258 (10)(2012) 4563–4568, **doi:10.1016/j.apsusc.2012.01.028**

[2] J.-C. Charlier, X. Blase, S. Roche. Electronic and transport properties of nanotubes. Rev. Mod. Phys. 79 (2)(2007) 677–732, **doi: 10.1103/RevModPhys.79.677**

[3] D. Tasis, N. Tagmatarchis, A. Bianco, M. Prato. Chemistry of Carbon Nanotubes. Chem. Rev. 106 (3)(2006) 1105−1136, **doi: 10.1021/cr050569o**

[4] R. H. Baughman, A. A.Zakhidov, W. A. de Heer. Carbon nanotubes - the route toward applications.Science 297 (5582)(2002) 787-792, **doi: 10.1126/science.1060928**

[5] M. S.Dresselhaus, G. Dresselhaus, P. Avouris. Carbon Nanotubes: Synthesis, Properties and Applications. 1st ed. Springer-Verlag Berlin Heidelberg(2001), **doi: 10.1007/3-540-39947-X**

[6] C. Li, T.-W. Chou.Elastic moduli of multi-walled carbon nanotubes and the effect of van der Waals forces. Compos. Sci. Technol. 63 (11)(2003) 1517–1524, **doi: 10.1016/S0266-3538(03)00072-1**

[7] J. P. Lu. Elastic properties of carbon nanotubes and nanoropes. Phys. Rev. Lett. 79 (7)(1997) 1297-1300, **doi: 10.1103/PhysRevLett.79.1297**

[8] A. Sobolkina, V. Mechtcherine, V. Khavrus, D. Maier, M. Mende, M. Ritschel, et al. Dispersion of carbon nanotubes and its influence on the mechanical properties of the cement matrix.Cem. Concr. Compos. 34 (10) (2012) 1104–1113,**doi: 10.1016/j.cemconcomp.2012.07.008**

[9] S. Boncel, S. W. Pattinson, V. Geiser, M. S. P. Shaffer, K. K. K. Koziol.En route to controlled catalytic CVD synthesis of densely packed and vertically aligned nitrogen-doped carbon nanotube arrays. Beilstein J. Nanotechnol. 5 (2014) 219–233, **doi: 10.3762/bjnano.5.24**

[10] R. Fuge, M. Liebscher, C. Schröfl, S. Oswald, A. Leonhardt, B. Büchner, et al. Fragmentation characteristics of undoped and nitrogen-doped multiwalled carbon nanotubes in aqueous dispersion in dependence on the ultrasonication parameters.Diamond Relat. Mater. 66 (2016) 126-134, **doi: 10.1016/j.diamond.2016.03.026**

[11] T. Sharifi, F. Nitze, H. R. Barzegar, C.-W. Tai, M. Mazurkiewicz, A. Malolepszy, et al. Nitrogen doped multi walled carbon nanotubes produced by CVD-correlating XPS and Raman spectroscopy for the study of nitrogen inclusion. Carbon 50 (10) (2012) 3535-3541, **doi: 10.1016/j.carbon.2012.03.022**

[12] K. K. K. Koziol, C. Ducati, A. H. Windle. Carbon Nanotubes with Catalyst Controlled Chiral Angle. Chem. Mater. 22 (17)(2010) 4904–4911, **doi: 10.1021/cm100916m**

[13] S. W. Pattinson, V. Ranganathan, H. K. Murakami, K. K. K. Koziol, A. H. Windle. Nitrogen-Induced Catalyst Restructuring for Epitaxial Growth of Multiwalled Carbon Nanotubes. ACSnano 6(9)(2012) 7723–7730, **doi: 10.1021/nn301517g**

[14] C. P. Ewels, M. Glerup. Review of Nitrogen Doping in Carbon Nanotubes. J. Nanosci. Nanotechnol. 5 (9) (2005) 1345–1365, **doi: 10.1166/jnn.2005.304**

[15] E. Cruz-Silva, D. A. Cullen, L. Gu, J. M. Romo-Herrera, E. Muñoz-Sandoval, F. López-Urías, et al. Heterodoped Nanotubes: Theory, Synthesis, and Characterization of Phosphorus-Nitrogen Doped Multiwalled Carbon Nanotubes. ACSnano 2(3)(2008) 441–448, **doi: 10.1021/nn700330w**

[16] E. N. Nxumalo, N. J. Coville. Nitrogen Doped Carbon Nanotubes from Organometallic Compounds: A Review. Materials 3 (3) (2010) 2141-2171, **doi:10.3390/ma3032141**

[17] A. Sobolkina, V. Mechtcherine, C. Bellmann, V. Khavrus, S. Oswald, S. Hampel, et al. Surface properties of CNTs and their interaction with silica. J. Colloid Interface Sci. 413(2014) 43–53, **doi: 10.1016/j.jcis.2013.09.033**

[18] J. Zhao, H. Lai, Z. Lyu, Y. Jiang, K. Xie, X. Wang, et al. Hydrophilic Hierarchical Nitrogen-Doped Carbon Nanocages for Ultrahigh Supercapacitive Performance. Adv. Mater. 27 (23) (2015) 3541–3545, **doi: 10.1002/adma.201500945**

[19] S. Ci, Z. Wen, J. Chen, Z. He. Decorating anode with bamboo-like nitrogen-doped carbon nanotubes for microbial fuel cells. Electrochem. Commun. 14 (1) (2012) 71–74, **doi: 10.1016/j.elecom.2011.11.006**

[20] R. Chetty, S. Kundu, W. Xia, M. Bron, W. Schuhmann, V. Chirila, et al. PtRu nanoparticles supported on nitrogen-doped multiwalled carbon nanotubes as catalyst for methanol electrooxidation. ElectrochimicaActa 54 (17)(2009) 4208–4215, **doi: 10.1016/j.electacta.2009.02.073**

[21] M. L. Zhao, D. J. Li, L. Yuan, Y. C. Yue, H. Liu, X. Sun. Differences in cytocompatibility and hemocompatibility between carbon nanotubes and nitrogen-doped carbon nanotubes. Carbon 49 (2011) 3125-3133, **doi: 10.1016/j.carbon.2011.03.037**

[22] E. Munoz-Sandoval, A. J. Cortes-López, B. Flores-Gómez, J. I. Fajardo-Díaz, R. Sánchez-Salas, F. López-Urías. Carbon sponge-type nanostructures based on coaxial

nitrogen-doped multiwalled carbon nanotubes grown by CVD using benzylamine as precursor.Carbon 115 (2017) 409-421, **doi: 10.1016/j.carbon.2017.01.010**

[23] S. Boncel, K. H. Müller, J. N. Skepper, K. Z. Walczak, K. K. K. Koziol.Tunable chemistry and morphology of multi-wall carbon nanotubes as a route to non-toxic,theranostic systems.Biomaterials 32 (30) (2011) 7677-7686, **doi: 10.1016/j.biomaterials.2011.06.055**

[24] H. Z. Wang, Z. P. Huang, Q. J. Cai, K. Kulkarni, C.-L. Chen, D. Carnahan, et al.Reversible transformation of hydrophobicity and hydrophilicity of aligned carbon nanotube arrays and buckypapers by dry processes.Carbon 48 (2010) 868-875, **doi: 10.1016/j.carbon.2009.10.041**

[25] V. Eckert, A. Leonhardt, S. Hampel, B. Büchner.Morphology of MWCNT in dependence on N-doping, synthesized using a sublimation-based CVD method at 750 °C.Diamond Relat. Mater. 86 (2018) 8-14, **doi: 10.1016/j.diamond.2018.04.004**

[26] K. Momma, F. Izumi. VESTA 3 for three-dimensional visualization of crystal, volumetric and morphology data. J. Appl. Crystallogr. 44 (2011) 1272-1276, **doi: 10.1107/S0021889811038970**

[27] N. R. Hore, D. K. Russell. Radical pathways in the thermal decomposition of pyridine and diazines: a laser pyrolysis and semi-empirical study. J. Chem. Soc., Perkin Trans. 2(1998) 269-275, **doi: 10.1039/A706731C**

[28] W. D. Crow, C. Wentrup. Reactions of excited molecules v. thermal decomposition of pyrazine. Tetrahedron Lett. 27(1968) 3115-3118, **doi: 10.1016/S0040-4039(00)89566-0**

[29] P. F. Britt. Pyrolysis and Combustion of Acetonitrile ($CH_3CN$). Oak Ridge National Laboratory, ORNL/TM-2002/113, **https://info.ornl.gov/sites/publications/Files/Pub57226.pdf**

[30] S. Song, D. M. Golden, R. K. Hanson, C. T. Bowman.A Shock Tube Study of Benzylamine Decomposition: Overall Rate Coefficient and Heat of Formation of the Benzyl Radical. J. Phys. Chem. A 106 (25)(2002) 6094-6098, **doi: 10.1021/jp020085l**

[31] M. He, S. Zhou, J. Zhang, Z. Liu, C. Robinson. CVD Growth of N-Doped Carbon Nanotubes on Silicon Substrates and Its Mechanism. J. Phys. Chem. B 109 (19)(2005) 9275-9279, **doi: 10.1021/jp044868d**

[32] H. C. Choi, J. Park, B. Kim. Distribution and Structure of N Atoms in Multiwalled Carbon Nanotubes Using Variable-Energy X-Ray Photoelectron Spectroscopy. J. Phys. Chem. B 109 (10)(2005) 4333-4340, **doi: 10.1021/jp0453109**

[33] I. Zeferino González, A. M. Valenzuela-Muñiz, G. Alonso-Nuñez, M. H. Farías, Y. Verde Gómeza. Influence of the Synthesis Parameters in Carbon Nanotubes Doped with Nitrogen for Oxygen Electroreduction. ECS J. Solid State Sci. Technol. 6 (6) (2017) 3135-3139, **doi: 10.1149/2.0251706jss**

[34] W.Y. Wonga, W.R.W. Daud, A.B. Mohamad, A.A.H. Kadhum, K.S. Loh, E.H. Majlan. Influence of nitrogen doping on carbon nanotubes towards the structure, composition and oxygen reduction reaction. Int. J. Hydrog. Energy 38 (22)(2013) 9421-9430, **doi: 10.1016/j.ijhydene.2013.01.189**

[35] A. Goldoni, R. Larciprete, L. Petaccia, S. Lizzit. Single-Wall Carbon Nanotube Interaction with Gases: Sample Contaminants and Environmental Monitoring. J. Am. Chem. Soc. 125 (37)(2003) 11329-11333, **doi: 10.1021/ja034898e**

[36] P. J. Schmitz, R. J. Baird. NO and $NO_2$ Adsorption on Barium Oxide: Model Study of the Trapping Stage of $NO_x$ Conversion via Lean $NO_x$ Traps. J. Phys. Chem. B 106 (16)(2002) 4172-4180, **doi: 10.1021/jp0133992**

[37] H. C. Choi, S. Y. Bae, J. Park, K. Seo, C. Kim, B. Kim, et al. Experimental and theoretical studies on the structure of N-doped carbon nanotubes: Possibility of intercalated molecular $N_2$. Appl. Phys. Lett. 85 (2004) 5742-5744, **doi: 10.1063/1.1835994**

[38] M. Glerup, V. Krstić, C. Ewels, M. Holzinger, G. Van Lier. Doping of Carbon Nanotubes. In: W. Chen, editor. Doped Nanomaterials and Nanodevices. American Scientific Publishers (2007), ISBN 1-58883-110-8

[39] J. Zhao, A. Buldum, J. Han, J. Ping Lu. Gas molecule adsorption in carbon nanotubes and nanotube bundles. Nanotechnology 13 (2002) 195-200, **doi: 10.1088/0957-4484/13/2/312**

[40] M. Seydou, S. Marsaudon, J. Buchoux, J. P. Aimé, A. M. Bonnot. Molecular mechanics investigation of carbon nanotube and graphene sheet interaction. Phys. Rev. B 80 (2009) 245421-245428, **doi: 10.1103/PhysRevB.80.245421**